\documentclass[english,a4paper,pre,twocolumn,amsmath,amssymb,superscriptaddress]{revtex4-1}
\usepackage{graphicx,color,soul}

\newcommand{\dd}{\mathrm{d}}

\newcommand{\mean}[1]{\langle #1 \rangle}

\newcommand{\IInt}[3]{\int_{#2}^{#3}\dd #1\;}
\newcommand{\unit}[1]{\;\mathrm{#1}}

\renewcommand{\vec}[1]{\mathbf #1}

\newcommand{\al}{\alpha}
\newcommand{\gam}{\gamma}
\newcommand{\eps}{\varepsilon}
\newcommand{\kap}{\kappa}
\newcommand{\lam}{\lambda}
\newcommand{\vhi}{\varphi}
\newcommand{\sig}{\sigma}

\newcommand{\id}{\mathbf 1}

\newcommand{\x}{\vec r}

\newcommand{\Dr}{D_\text{r}}

\newcommand{\nois}{\boldsymbol\xi}

\newcommand{\kT}{k_\text{B}T}

\begin{document}

\title{Aggregation and sedimentation of active Brownian particles at constant affinity}

\author{Andreas Fischer}
\affiliation{Institut f\"ur Physik, Johannes Gutenberg-Universit\"at Mainz, Staudingerweg 7-9, 55128 Mainz, Germany}
\affiliation{Graduate School Materials Science in Mainz, Staudinger Weg 9, 55128 Mainz, Germany}
\author{Arkya Chatterjee}
\affiliation{Department of Physics, Indian Institute of Technology-Bombay, Powai, Mumbai 400076, India}
\author{Thomas Speck}
\affiliation{Institut f\"ur Physik, Johannes Gutenberg-Universit\"at Mainz, Staudingerweg 7-9, 55128 Mainz, Germany}

\begin{abstract}
  We study the motility-induced phase separation of active particles driven through the interconversion of two chemical species controlled by ideal reservoirs (chemiostats). As a consequence, the propulsion speed is non-constant and depends on the actual inter-particle forces, enhancing the positive feedback between increased density and reduced motility that is responsible for the observed inhomogeneous density. For hard discs, we find that this effect is negligible and that the phase separation is controlled by the average propulsion speed. For soft particles and large propulsion speeds, however, we predict an observable impact on the collective behavior. We briefly comment on the reentrant behavior found for soft discs. Finally, we study the influence of non-constant propulsion on the sedimentation profile of non-interacting active particles.
\end{abstract}

\maketitle


\section{Introduction}

The defining characteristic of active particles is their directed motion, through which they exert mechanical forces on their environment~\cite{bech16}. A wealth of different mechanisms has been proposed to achieve such directed motion of synthetic colloidal particles~\cite{dey17}, most prominently self-phoresis~\cite{paxt04,hows07,pala10,butt13}, but also the scattering of sound waves~\cite{wang12} and particles driven by an external electric field~\cite{bric13,yan16a}. Ignoring phoretic and hydrodynamic interactions, active Brownian particles (ABPs) have become a widely studied model system combining the directed motion of single particles with short-ranged repulsive interactions between particles~\cite{yaou12,redn13,bial13,wyso14,solon15,levi17}. Strikingly, even in the absence of cohesive forces one observes a dynamical instability through which the system becomes inhomogeneous, which has been termed motility-induced phase separation (MIPS)~\cite{bial14,cate15}. The superficial similarity of MIPS with passive liquid-gas separation has lead to mappings onto effective dynamics obeying detailed balance with isotropic attractions~\cite{fara15,rein16a}. However, the following observations preclude the validity of such an effective equilibrium picture for ABPs: (i)~the phase equilibrium is not determined by the Maxwell construction~\cite{solo17,paliwal18}, (ii)~the interfacial tension is not given by the interfacial fluctuations and, moreover, negative~\cite{bial15}, and (iii)~fluctuations close to the critical point seem to be different from Ising universality~\cite{sieb18}.

Being established as a genuine non-equilibrium system, recently the entropy production of active Brownian particles has received attention. Most works have focused on the \emph{trajectory entropy production} following from the probability ratio of forward to backward trajectories. A departure of this ratio from unity indicates breaking of time-reversal symmetry, and thus non-equilibrium. In stochastic thermodynamics~\cite{seif12}, this trajectory entropy production equals the heat following from the first law. Hence, we need to identify unambiguously the energy being exchanged as heat with the environment, a route that is not available for ABPs from the model equations of motion alone without further assumptions. This lack of physical detail is the deeper reason for the recent multitude of conflicting results for the (apparent) entropy production of ABPs~\cite{gang13,fodo16,puglisi17,mandal17,marc17,shankar18,mandal18,caprini18}.

One way to cure this shortcoming is to explicitly model the mechanism leading to the directed motion of particles. The arguably simplest model is the reaction of an abundant chemical that induces a local gradient propelling the particle along its orientation, either treating the concentration of the chemical explicitly~\cite{yan16,huang17} or assuming chemiostats~\cite{piet17,spec18}. The latter approach ignores depletion and inhomogeneities of the chemical species, and therefore phoretic interactions that would be mediated by these concentration gradients~\cite{pohl14,lieb15}, and leads to a simple modification of ABPs in which the propulsion speed of each particle depends on the conservative forces exerted by neighboring particles. In the following, we study the consequences of this generalized ABPs model for the motility-induced phase separation. Moreover, we analytically calculate the sedimentation length of non-interacting active particles~\cite{tail09,pala10,encu11,gino15,hermann18} under the influence of a constant force.


\section{Model}

\begin{figure}[t]
  \centering
  \includegraphics{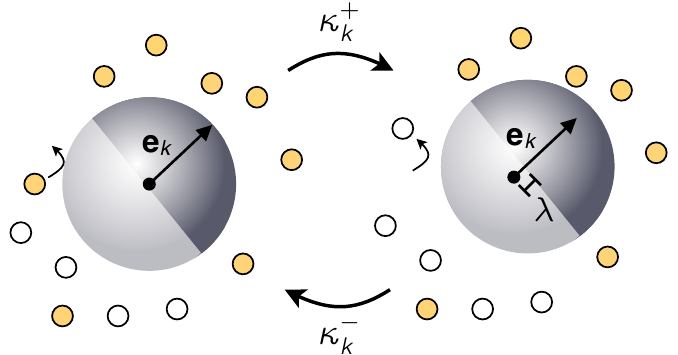}
  \caption{Modeling the self-propulsion of a Janus particle. In response to each chemical conversion $\bullet\to\circ$ (see arrow), the particle is translated along its orientation $\vec e_k$ by the distance $\lam$. The corresponding rate is $\kap^+_k$, whereas the (non-zero) rate for a step back is $\kap^-_k$. Both rates obey the detailed balance condition.}
  \label{fig:chemical}
\end{figure}

We consider a suspension of $N$ active particles that undergo directed motion through the conversion of free energy. We assume that part of the particle surface acts as catalyst for the chemical reaction $\bullet\leftrightharpoons\circ$ between two generic chemical species, which only occurs on the surface and the solvent acts as a particle (and heat) reservoir. This reaction could explicitly model the decomposition of hydrogen peroxide, but also the conversion from lutidine-rich to lutidine-poor in the case of diffusiophoresis. For conceptual simplicity, we assume the corresponding chemical potentials to be fixed, which implies ideal particle reservoirs at constant concentrations (chemiostats)~\cite{seif11}. The difference $\Delta\mu$ of the chemical potentials is also called the \emph{driving affinity}. In response to every chemical event, particles are displaced by the length $\lam$ along their orientations $\vec e_k$. This length depends on many factors such as the particle geometry and surface properties~\cite{gole07}, and is treated as a parameter in the following. The propulsion speed of particle $k$ is then non-constant and in the limit of small $\lam/\sig\ll1$ reads~\cite{spec18}
\begin{equation}
  \label{eq:v:def}
  \hat v_k = \lam(\kap^+_k-\kap^-_k)
\end{equation}
with rates $\kap^\pm_k$ for a step forward and backward, respectively, cf. Fig.~\ref{fig:chemical}. Here, $\sig$ is a length scale typically related to the particle size. If $\Delta\mu$ is positive, the reaction produces species $\circ$ on average and particles are propelled in the direction of $\vec e_k$. Note that incorporating more elaborate models for the chemical dynamics -- such as Michaelis-Menten kinetics -- instead of this simple tight-binding model is straightforward.

For the rates, we assume the symmetric form
\begin{equation}
  \label{eq:rates}
  \kap^\pm_k = \kap_0 e^{\pm\beta(\Delta\mu-\hat f_k)/2}
\end{equation}
with attempt rate $\kap_0$, which obey the detailed balance condition with respect to the potential energy $U(\{\x_k\})$. The surrounding solvent acts as heat reservoir at constant temperature $T$ with $\beta\equiv(\kT)^{-1}$. Here, $\hat f_k\equiv\lam\vec e_k\cdot\nabla_kU$ is the change of potential energy due to a (small) displacement of particle $k$. Plugging Eq.~(\ref{eq:rates}) into Eq.~(\ref{eq:v:def}), the individual speed of particle $k$ thus reads
\begin{equation}
  \label{eq:v}
  \hat v_k = 2\kap_0\lam\sinh[\beta(\Delta\mu-\hat f_k)/2].
\end{equation}
In the absence of a potential energy ($U=0$), the free speed becomes
\begin{equation}
  \label{eq:v0}
  v_0 \equiv 2\kap_0\lam\sinh(\beta\Delta\mu/2).
\end{equation}
We thus have the three parameters $\Delta\mu$, $\lam$, and $\kap_0$ that characterize the directed motion of the particles.

The coupled equations of motion read
\begin{equation}
  \label{eq:eom}
  \dot\x_k = \hat v_k\vec e_k - \mu_0\nabla_kU + \nois_k
\end{equation}
with translational noises having zero mean and correlations $\mean{\nois_k(t)\nois_l^T(s)}=2D_0\delta_{kl}\id\delta(t-s)$. The bare diffusion coefficient $D_0=\kT\mu_0$ obeys the Stokes-Einstein relation with bare mobility $\mu_0$. In addition, the orientations undergo free rotational diffusion with diffusion coefficient $\Dr$.


\section{Motility-induced phase separation}

\subsection{Effective speed}

It is now well established that ABPs exhibit a transition to an inhomogeneous state (at fixed area and particle number) with a density $\rho(\x)$ that varies spatially~\cite{cate15}. The central quantity for the theoretical description of this phenomenon is played by the effective speed
\begin{equation}
  \label{eq:veff}
  v(\rho) = \mean{\dot\x_k\cdot\vec e_k}_\rho =
  \mean{\hat v_k - \mu_0\vec e_k\cdot\nabla_kU}_\rho,
\end{equation}
where the average is over an approximately homogeneous region of the system with density $\rho$. Again appealing to $\lam/\sig\ll1$, for small $\beta\hat{f}_k\ll 1$, we expand Eq.~\eqref{eq:v} yielding the approximation
\begin{equation}
  \label{eq:v:app}
  \hat v_k \approx v_0 - \chi\mu_0\vec e_k\cdot\nabla_k U
\end{equation}
with dimensionless coefficient
\begin{equation}
  \label{eq:chi}
  \chi \equiv \frac{\kap_0 \lam^2}{D_0}\sqrt{1+\left(\frac{v_0}{2\kappa_0\lambda}\right)^2}.
\end{equation}
For growing $\beta\Delta\mu>1$, this coefficient quickly approaches $\chi\approx v_0\lam/(2D_0)$.

Plugging Eq.~\eqref{eq:v:app} into Eq.~\eqref{eq:veff}, we obtain the effective speed
\begin{equation}
  \label{eq:veff:app}
  v(\rho) = v_0 - (1+\chi)\mu_0\mean{\vec e_k\cdot\nabla_k U}_\rho.
\end{equation}
For pure ABPs driven by constant speed $v_0$, we find the same expression with $\chi=0$~\cite{bial13}, where the effective speed is reduced due to the blocking of particles. Hence, a non-constant propulsion speed with $\chi>0$ further reduces the effective speed. In the following, we investigate how this reduction affects the phase behavior for two specific interaction potentials.

\subsection{Hard discs}

\begin{figure*}[t]
  \centering
  \includegraphics[scale=.9]{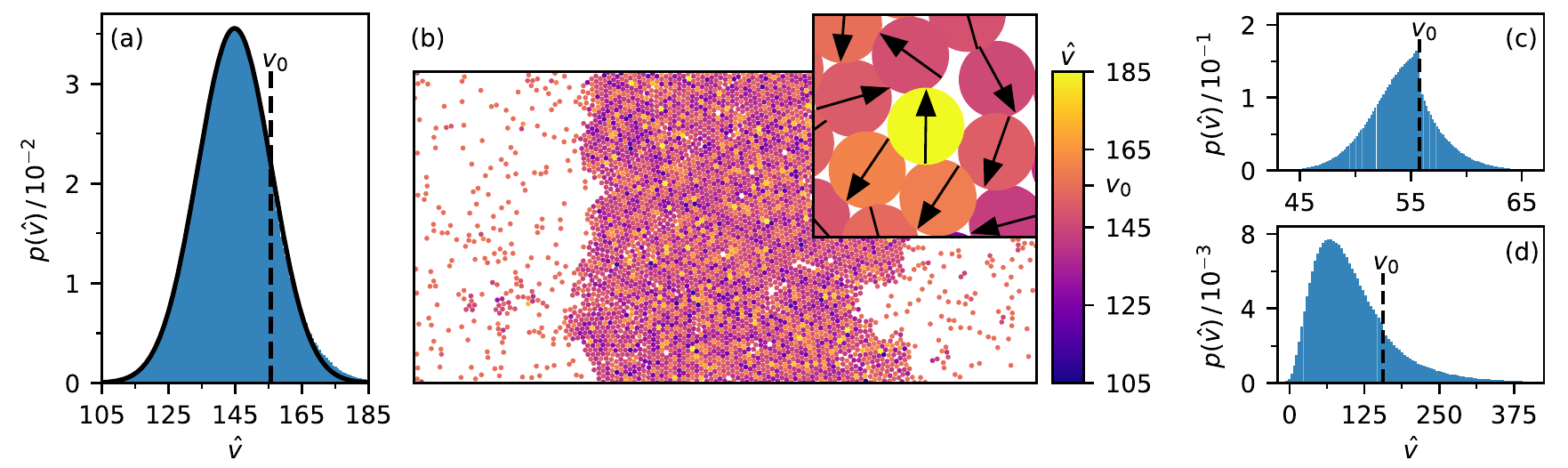}
  \caption{(a)~Distribution of speeds $\hat v_k$ of interacting particles. Indicated is the speed $v_0$ of free particles, which are not included in the distribution. The solid line is a fitted Gaussian with mean $\mean{\hat v}\simeq 145$ and standard deviation $\simeq 11$. (b)~Simulation snapshot showing the dense slab surrounded by an active gas. The color indicates the instantaneous propulsion speed $\hat v_k$. The inset shows a particle configuration in the dense phase. The tagged particle (yellow) at the center has an increased speed $\hat{v}_k>v_0$ as it interacts with three particles at its back and only one at its front, implying $\vec e_k\cdot\nabla_k U<0$. (c,d)~Non-Gaussian speed distributions for (c)~$\lam=10^{-3}$, $\Delta\mu=3.5$ and (d)~$\lam=10^{-2}$, $\Delta\mu=5.5$.}
  \label{fig:wca}
\end{figure*}

We now study suspensions of interacting particles, where the potential energy $U=\sum_{k<l}u(|\x_k-\x_l|)$ stems from pairwise interactions with pair potential $u(r)$ depending on the particle separation. We first consider the Weeks-Chandler-Andersen (WCA) potential~\cite{week72}
\begin{equation}
  \label{eq:wca}
  u(r) =
  \begin{cases}
    4\eps\left[\left(\frac{\sig}{r}\right)^{12} - \left(\frac{\sig}{r}\right)^6 
      + \frac{1}{4}\right] & (r/\sig < 2^{1/6}) \\
    0 & (r/\sig \geqslant 2^{1/6}).
  \end{cases}
\end{equation}
We employ parameters as in previous studies~\cite{bial15,sieb17,sieb18}: The rotational diffusion coefficient is set to $\Dr=3D_0/\sigma_\text{eff}^2$ and $\epsilon=100\kT$, which implies harsh repulsion and models (almost) hard discs with an effective diameter $\sig_\text{eff}=1.10688\sig$. In this section, we employ $\sigma$, $\sigma^2/D_0$, and $\kT$ as units of length, time, and energy, respectively. Throughout, we fix the product $\kap_0\lam=10$ and vary the free speed $v_0$ by changing $\Delta \mu$ for several lengths $\lam$. We perform Brownian dynamics simulations integrating Eq.~\eqref{eq:eom} with time step $\Delta t=2\cdot 10^{-6}$. The implementation of the non-constant propulsion speed is straightforward as it involves the total force on each particle. The simulations are carried out in an elongated box (size ratio 2:1) with periodic boundary conditions at global density $\bar{\rho}\simeq0.52$ and sufficiently high speeds $v_0$ so that the system separates into a dense slab and a dilute gas~\cite{bial15,sieb17,sieb18}, cf. Fig.~\ref{fig:wca}(b).

While most particles in the active gas move with the free speed $v_0$, in the dense phase the individual speeds $\hat v_k$ acquire a broad distribution, see Fig.~\ref{fig:wca}(a)
for $\lam=10^{-3}$ and $\Delta\mu=5.5$. As one would expect, mostly the speeds are smaller than the free speed $v_0$, in particular the mean is smaller. However, there is a tail towards larger speeds. When the total force on a particle points along its orientation, its propulsion speed can rise above $v_0$. This occurs, \emph{e.g.}, for particles interacting with more particles in its back than its front [see inset of Fig.~\ref{fig:wca}(b)], so that the directed step can actually lower the potential energy of the system. Fig.~\ref{fig:wca}(b) suggests that different speeds $\hat{v}_k$ are distributed rather uniformly throughout the dense phase without long-ranged correlations. Closer to the critical point (\emph{i.e.}, reducing $\Delta \mu$), the speed distribution becomes markedly non-Gaussian, cf. Fig.~\ref{fig:wca}(c). Fig.~\ref{fig:wca}(d) shows that at higher $\lam=10^{-2}$ the speed distribution becomes broader (due to the higher sensitivity of the speed on the forces) and exhibits a notably asymmetric shape leaning towards large speeds.

\begin{figure*}[t]
  \centering
  \includegraphics[scale=.9]{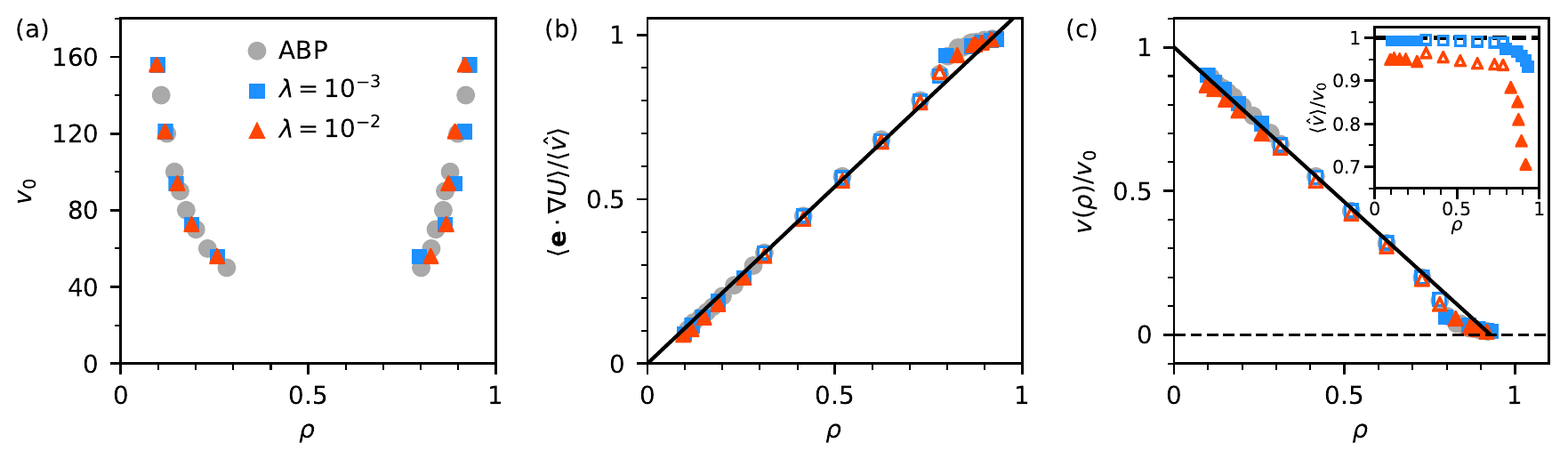}
  \caption{Phase behaviour of hard discs. (a)~Phase diagram (coexisting densities) for ABPs with constant speed $v_0$ (data taken from Ref.~\cite{sieb17}) and for ABPs at constant affinity for different values of $\lambda$. (b)~Average force opposing the propulsion $\mean{\vec e_k \cdot \nabla_k U}_\rho$ as a function of density [same color code as in (a)]. Open symbols show data acquired in the homogeneous regime at $v_0=20$. Black line: Fit $a\rho$ to all data points up to $\rho=0.7$ with $a\simeq 1.1$. (c)~Effective speed $v(\rho)$ as defined in Eq.~\eqref{eq:veff} and normalized by $v_0$. The black line is Eq.~\eqref{eq:vMean_wca} for ABPs at constant propulsion speed ($\chi=0$) with $a$ as obtained in (b). Inset: speed $\mean{\hat{v}_k}_\rho/v_0$ for ABPs at constant affinity. Larger $\lam$ and larger density $\rho$ cause a stronger reduction of the speed.}
  \label{fig:wca2}
\end{figure*}

The slab geometry encourages the interface between gas and dense phase to align with the shorter box edge. Employing discrete bins along the elongated edge of the box, this allows us to determine the coexisting densities in the phase-separated regime from the plateau values of the time-averaged density distribution (with the center-of-mass shifted to zero before time averaging), see Refs.~\cite{bial15,sieb17,sieb18}. Since the slab has a straight interface [see Fig.~\ref{fig:wca}(b)], this method minimizes finite-size effects. By fitting both sides of the density profile using
\begin{equation}
\label{eq:tanhFit}
 \rho(x)=\frac{\rho_\text{liq}+\rho_\text{gas}}{2}+\frac{\rho_\text{liq}-\rho_\text{gas}}{2}\tanh\left(\frac{x-x_0}{2\omega}\right)
\end{equation}
we extract the coexisting densities $\rho_\text{gas}$ and $\rho_\text{liq}$ together with the width of the interface $\omega$. In Fig.~\ref{fig:wca2}(a), we plot the coexisting densities for pure ABPs (constant speed $v_0$) and ABPs at constant affinity [with $v_0$ given through Eq.~\eqref{eq:v0}] for $\lambda=10^{-3}$ and $10^{-2}$, which lie on top of each other.

In order to obtain further insight, we also spatially resolve the average propulsion speed $\mean{\hat{v}_k}_\rho$ and the average force opposing the propulsion direction $\left<\vec{e}_k\cdot\nabla_k U\right>_\rho$ in the two phases, which show a similar behavior as the density profile. We find that for larger $\lam$, $\mean{\hat{v}_k}_\rho$ is reduced stronger as can be seen by comparing the speed distributions in Fig.~\ref{fig:wca}(a) and (d). As $\mean{\hat{v}_k}_\rho$ decreases, the frequency of particle collisions also decreases, leading to a simultaneous reduction of
$\mean{\vec{e}_k\cdot \nabla_k U}_\rho$. Thus $\mean{\vec{e}_k\cdot\nabla_k U}_\rho\propto\mean{\hat{v}_k}_\rho$, which is confirmed by the data shown in Fig.~\ref{fig:wca2}(b). For densities up to $\rho\lesssim 0.7$, we find 
\begin{equation}
  \label{eq:Fv_lin}
  \mean{\vec{e}_k\cdot \nabla_k U}_\rho\approx a \rho \mean{\hat{v}_k}_\rho
\end{equation}
with $a\simeq 1.1$ in agreement with Ref.~\cite{sten13}. At low densities, the data exhibit small deviations from the linear behavior. These data points correspond to the dilute phase at high propulsion speeds and possibly suffer from finite-size effects. For high densities, the ratio $\mean{\vec{e}_k\cdot\nabla_k U}_\rho/\mean{\hat{v}_k}_\rho$ no longer scales proportional to the density and eventually saturates at $\simeq 1$, indicating arrest in the dense phase. In addition, we have performed simulations at lower speeds in the homogeneous phase. The resulting data is plotted in Fig.~\ref{fig:wca2} using open symbols and fits perfectly with the data from the inhomogeneous systems.

In Fig.~\ref{fig:wca2}(c), we plot the effective speed $v(\rho)$ normalized by the free speed $v_0$, which collapses onto an almost linear master curve (there is a small systematic dependence on $\lam$). In contrast, the average propulsion speed $\mean{\hat{v}_k}_\rho$ [inset of Fig.~\ref{fig:wca2}(c)] is affected by changing $\lam$ and is reduced at higher densities. However, this effect is approximately compensated by $\mean{\vec{e}_k\cdot \nabla_k U}_\rho$ according to Eq.~\eqref{eq:Fv_lin}. Combining Eq.~\eqref{eq:Fv_lin} with the average of Eq.~\eqref{eq:v:app}, and plugging the result into Eq.~\eqref{eq:veff:app}, we arrive at the effective speed
\begin{equation}
  \label{eq:vMean_wca}
  v_\text{WCA}(\rho) = v_0\left[1-\frac{a (1+\chi)}{1+a\rho\chi}\rho\right],
\end{equation}
which comprises the effects discussed above. Note that $\chi$ depends on $v_0$ [Eq.~\eqref{eq:chi}].

\subsection{Dynamic mean-field theory}
\label{sec:meanfield}

Following Ref.~\cite{bial13}, we derive a dynamical equation for the one-body density $\psi(\vec{r},\vhi,t)$ with angle $\vhi$ enclosed by the orientation with fixed $x$-axis. The starting point is the full many-body Smoluchowski equation for the joint probability of all particle positions and orientations. Employing the expansion Eq.~\eqref{eq:v:app} and integrating over positions and orientations except for a tagged particle with $(\x_k,\vhi_k)$ yields
\begin{equation}
  \partial_t\psi = -\nabla\cdot[v(\rho)\vec e-\nabla]\psi 
  + \Dr\partial_\vhi^2\psi
\end{equation}
with effective speed $v(\rho)$ defined in Eq.~\eqref{eq:veff:app}, where we drop the indices.

The two lowest moments with respect to angle $\vhi$ are the density
\begin{equation}
  \rho(\x,t) \equiv \IInt{\vhi}{0}{2\pi} \psi(\x,\vhi,t)
\end{equation}
and the polarization
\begin{equation}
  \vec{p}(\x,t) \equiv \IInt{\vhi}{0}{2\pi} \vec{e} \psi(\x,\vhi,t),
\end{equation}
which obey the coupled dynamic equations
\begin{gather}
  \label{eq:mf_dens}
  \partial_t \rho =-\nabla \cdot [v(\rho)\vec p]+\nabla^2\rho, \\
  \label{eq:mf_pol}
  \partial_t \vec p =-\frac{1}{2} \nabla [v(\rho)\rho]+\nabla^2 \vec p-\Dr \vec p. 
\end{gather}
Performing a linear stability analysis of Eqs.~\eqref{eq:mf_dens} and \eqref{eq:mf_pol} with respect to the homogeneous state $\rho(\x,t)=\bar{\rho}=\mathrm{const.}$, $\vec{p}(\x,t)=0$, one finds that perturbations at wave vector $q$ grow with rate
\begin{equation}
  \sig(q) = -\left(1+\frac{\al \gam}{\Dr}\right)q^2-(\al\gam)^2q^4 +\mathcal{O}(q^6).
\end{equation}
For $\Dr+\al \gam<0$ perturbations with small $q$ grow, leading to a dynamical instability. Plugging in Eq.~\eqref{eq:vMean_wca} for the hard discs, the coefficients read $\al=v_\text{WCA}(\bar{\rho})$ and
\begin{equation}
  \label{eq:gam}
  \begin{split}
    \gam &= 
    \frac{1}{2} \left(v_\text{WCA}(\bar{\rho})+\bar{\rho} \frac{\partial v_\text{WCA}(\rho)}{\partial \rho} \bigg|_{\bar{\rho}} \right) \\
    &= \frac{1}{2}v_0 - \frac{v_0a(1+\chi)}{1+a\bar\rho\chi}\bar\rho + \frac{1}{2}\frac{v_0a^2(1+\chi)\chi}{(1+a\bar\rho\chi)^2}\bar\rho^2.
  \end{split}
\end{equation}
In the case of ABPs at constant speed with $\chi=0$, the instability condition can be easily solved analytically for the instability line (spinodal)~\cite{spec15}
\begin{equation}
  \label{eq:rho:sp}
  \rho_\pm = \frac{3}{4a} \pm \frac{1}{4a}\sqrt{1-\frac{v_\text{c,ABP}^2}{v_0^2}}
\end{equation}
implying a critical point at speed $v_\text{c,ABP}=4\sqrt{\Dr}$ and density $\bar{\rho}_\text{c,ABP}=3/(4a)$ with $a$ from Eq.~\eqref{eq:Fv_lin}. 

\begin{figure}[t]
  \centering
  \includegraphics[scale=.9]{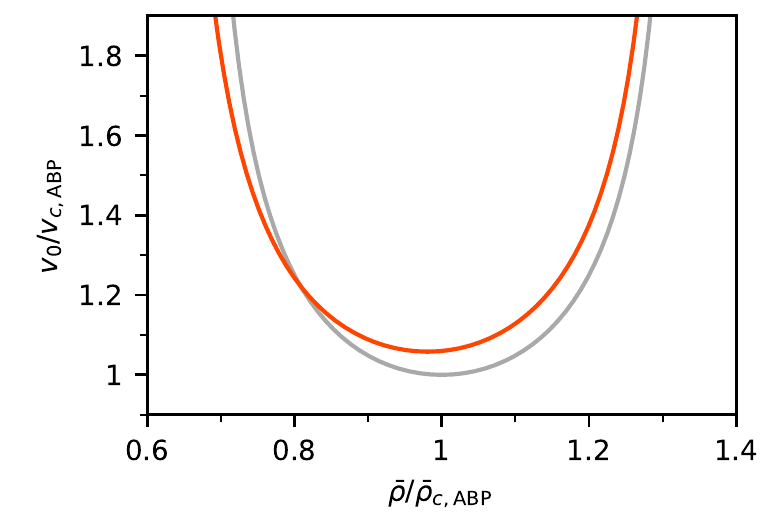}
  \caption{Mean-field spinodals for pure ABPs (gray) and ABPs at constant affinity for $\lam=10^{-2}$ (red).}
  \label{fig:spinodal}
\end{figure}

In order to determine the effect of $\chi>0$, we scale the density $\bar\rho=\bar\rho_\text{c,ABP}(1+x)$ and expand the coefficients to linear order of $x$. This leads to a quadratic equation for $x$ (solution in appendix~\ref{sec:spin}). In the limit of vanishing $\chi\to0$, we recover Eq.~\eqref{eq:rho:sp}. Employing Eq.~\eqref{eq:chi} for $\chi$ with $\kap_0\lam=10$ and varying $v_0$ at fixed $\lam$, we plot the mean-field spinodals in Fig.~\ref{fig:spinodal} for pure ABPs and $\lam=10^{-2}$. We see that for non-constant propulsion speeds, the critical point shifts to lower densities, $\bar\rho_\text{c}<\bar\rho_\text{c,ABP}$, but larger speeds $v_\text{c}>v_\text{c,ABP}$. In the actual simulations of hard discs, this shift appears to be negligible.

\subsection{Soft discs}

In a second system, we investigate the impact of the softness of the interaction potential on the collective behavior. To this end, we perform simulations with the harmonic interaction potential
\begin{equation}
  \label{eq:hp}
  u(r) =
  \begin{cases}
    \eps(r/\sig-1)^2 & (r/\sig < 1) \\
    0 & (r/\sig \geqslant 1),
  \end{cases}
\end{equation}
where now particles can overlap at a finite energy cost. As in Ref.~\cite{bial13}, we employ $\eps=100$, a reduced rotational diffusion coefficient $\Dr=3\cdot 10^{-3}$, and gather data at global density $\bar{\rho}\simeq0.89$. As for the hard discs, we keep the product $\kap_0\lam=10$ constant and adjust $\Delta\mu$ for given $v_0$. For ABPs with constant propulsion speed, we observe phase separation at speeds below $v_0\approx15$ with a stable slab forming for $v_0 \gtrsim 7.5$. For $v_0 \gtrsim 15$, we observe a reentrance into the disordered phase as previously reported in Ref.~\cite{bial13}.

\begin{figure}[t]
  \centering
  \includegraphics[scale=.9]{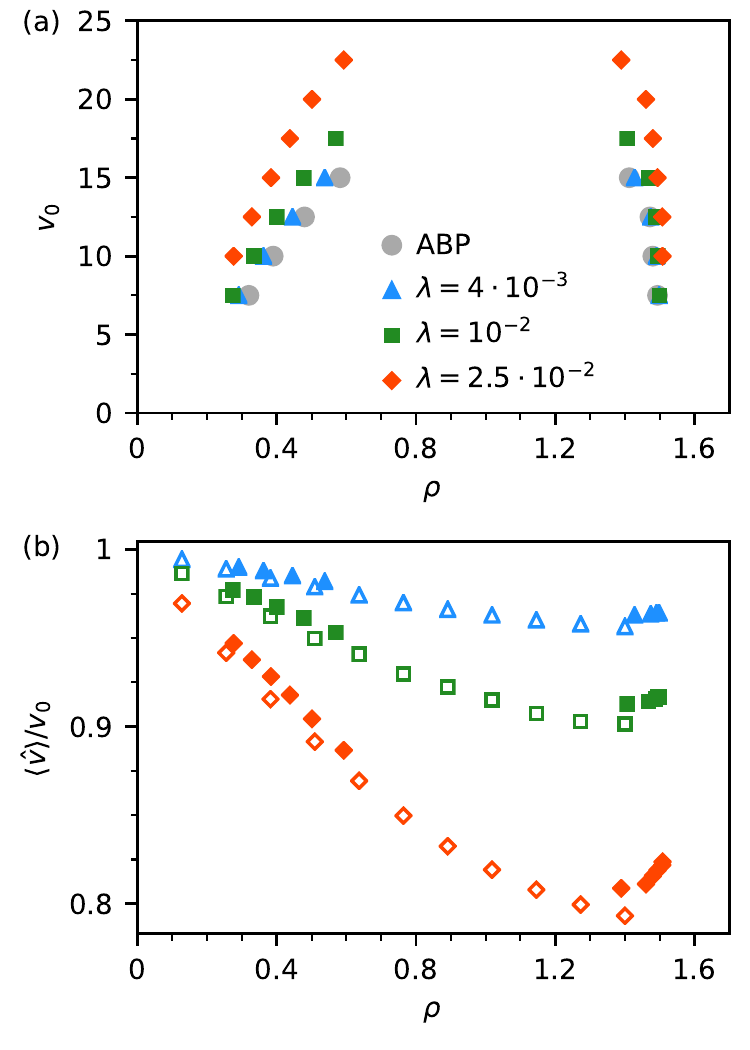}
  \caption{Phase behaviour of soft discs. (a)~Coexisting densities for ABPs with constant speed $v_0$ and for ABPs at constant affinity for different values of $\lambda$. (b)~Speed $\mean{\hat{v}_k}_\rho/v_0$ for ABPs at constant affinity [same color code as in (a)]. Open symbols show additional data acquired in the homogeneous regime at $v_0=30$. } 
  \label{fig:harm}
\end{figure}

In Fig.~\ref{fig:harm}(a), we plot the coexisting densities in the phase-separated regime. First, our data suggests a binodal that closes in an upper critical point in agreement with the reentrant behavior. Second, we now observe a much larger impact of the non-constant speed on the coexisting densities, with the binodal being shifted to larger speeds when increasing $\lam$ and the difference between the coexisting densities (at fixed $v_0$) being enhanced. In Fig.~\ref{fig:harm}(b), we show the propulsion speed $\mean{\hat v_k}_\rho$ for different $\lambda$ in the inhomogeneous system together with data in the homogeneous system at $v_0=30$. Here, the propulsion speed differs significantly from $v_0$ over a large range of densities. As in the case of hard discs, the deviation becomes more pronounced with increasing $\lambda$. In contrast to hard disks, however, already at constant speed the effective speed $v(\rho)$ does not collapse onto a single master curve anymore when normalized by $v_0$, but rather depends on $v_0$ in a more complicated manner (see appendix~\ref{sec:soft}). Thus, due the strong dependence of $\mean{\hat v_k}_\rho$ on $\lambda$, $v(\rho)$ is affected considerably when we change $\lam$, resulting in a shift of the binodal.

\section{Sedimentation}

Finally, we study the sedimentation of an ideal gas of active particles in the external potential $U(z)=mgz$ in a two-dimensional geometry. It is sufficient to consider the one-body density $\psi(z,\vhi,t)$ as a function of height $z\geqslant 0$, orientation $\vhi$ with respect to the $z$-axis, and time $t$, which evolves according to
\begin{equation}
  \label{eq:sed}
  \partial_t\psi = -(\hat v\cos\vhi-\mu_0mg)\partial_z\psi + D_0\partial_z^2\psi + \Dr\partial_\vhi^2\psi.
\end{equation}
We again employ the expansion Eq.~\eqref{eq:v:app} for the speed, which now yields $\hat v\approx v_0-\chi\mu_0mg\cos\vhi$. Following Ref.~\cite{hermann18}, in the steady state we insert the product ansatz $\psi(z,\vhi)=f_\xi(\vhi)e^{-z/\xi}$. Using $\cos^2\vhi=\tfrac{1}{2}(1+\cos2\vhi)$ and substituting $\vhi=2x$, we obtain the Whittaker-Hill ordinary differential equation
\begin{equation}
  \label{eq:whit-hill}
  0 = f_\xi''(x) + (A-2B\eps\cos 2x-2\eps^2\cos 4x)f_\xi(x)
\end{equation}
with coefficients
\begin{gather*}
  \eps^2 = \al\chi = \al\sqrt{1+X^2}\frac{\kap_0\lam^2}{D_0} \propto \lam^2 \\
  A=-4\al-2\eps^2+4D_0/(\Dr\xi^2) \\
  B = -\frac{2}{\Dr\xi}\left(\kap_0D_0\frac{X^2}{\al\sqrt{1+X^2}}\right)^{1/2}
\end{gather*}
defining $\al\equiv\mu_0mg/(\Dr\xi)$ and $X\equiv\sinh(\beta\Delta\mu/2)$. Note that $B$ is independent of $\lam$.

We now seek an analytical solution that is periodic, $f_\xi(x+\pi)=f_\xi(x)$. Eq.~\eqref{eq:whit-hill} can be interpreted as an eigenvalue equation for $A$, for which an expansion for small $\eps\ll 1$ (\emph{viz.} small $\lam$) yields 
\begin{equation}
  A = -\eps^2\frac{B^2}{2} + \mathcal O(\eps^4).
\end{equation}
Details can be found, \emph{e.g.}, in Ref.~\cite{rim15}. Inserting the expressions for the coefficients, we find
\begin{equation}
  \left(1+\frac{\chi}{2}\right)\frac{\xi}{\xi_\text{eq}} = 1+\frac{v_0^2}{2D_0\Dr}
\end{equation}
with equilibrium sedimentation length $\xi_\text{eq}\equiv D_0/(\mu_0mg)$. For pure ABPs with $\chi=0$, we thus recover the result of Ref.~\cite{hermann18} for the sedimentation length $\xi$. Including the fluctuations of the propulsion speed, the sedimentation length is reduced by the factor $(1+\chi/2)^{-1}\approx(1-\chi/2)$.


\section{Conclusions}

In a thermodynamically consistent model for interacting self-propelled colloidal particles driven by a chemical potential difference, the self-propulsion speed necessarily is non-constant and depends on potential energy (differences). A strictly constant speed would require an external agent controlling speeds individually~\cite{bauerle18}. The influence of this non-constant speed on the motility-induced phase separation and sedimentation length is captured by a single dimensionless quantity, $\chi$. In the simulations performed for hard discs, we have explored values for $\lam$ and $v_0$ that yield $\chi\sim10^{-2}\dots1$.

In experiments with colloidal particles, it is difficult to directly estimate the values for the displacement length $\lam$ and the attempt rate $\kap_0$. For not too small affinity $\Delta\mu$, Eq.~\eqref{eq:chi} reduces to $\chi\approx v_0\lam/(2D_0)$, where $\kap_0$ does not appear independently. For diffusiophoresis~\cite{butt13}, particles reach speeds on the order $v_0\sim1\unit{\mu m/s}$ and have diffusion coefficients $D_0\sim0.1\unit{\mu m^2/s}$ (for radius $R\simeq2\unit{\mu m}$), thus $\chi\sim5\lam\unit{\mu m}^{-1}$. If we assume that $\lam\sim1\unit{nm}$, we would have $\chi\sim10^{-3}$. Hence, for most practical purposes we can employ the standard model of ABPs with constant speed. Interestingly, this holds even though the fluctuations of the propulsion speed $\hat v$ can become quite large [cf. Fig.~\ref{fig:wca}(d)]. Even for a distribution of speeds due to, \emph{e.g.}, variations in the particle preparation, we thus expect that the collective behavior remains unaffected. However, we stress that conceptually the introduction of non-constant propulsion speeds is important when addressing thermodynamic questions like the entropy production.

Bacteria have sizes and diffusion coefficients that are comparable to colloidal particles but reach larger propulsion speeds (\emph{e. coli}: length $2\unit{\mu m}$ and $v_0\approx30\unit{\mu m/s}$), which thus yields larger $\chi$. Moreover, these objects are deformable and we have shown that for soft interactions the impact of a non-constant speed on the collective behavior is not negligible anymore.

Going to smaller scales, catalytic enzymes have been shown to undergo enhanced diffusion~\cite{sengupta13}, which has been characterized using the active Brownian particle model~\cite{jee18}. The chemical cycle of enzymes is typically described by Michaelis-Menten kinetics rather than the tight-binding model employed here. Nevertheless, plugging in numbers for the passive diffusion $D_0\sim 10^{3}\unit{\AA^2/\mu s}$ and the boost speed $v_0\sim 10\unit{\AA/\mu s}$, we find $\chi\sim10^{-2}\lam/\mathrm{\AA}$. Since the displacement length should be related to conformational changes of the enzyme, we assume that $\lam\lesssim\mathrm{\AA}$ is of the order of the enzyme size. Hence, $\chi\lesssim 10^{-2}$ is again very small due to the large translational diffusion coefficient that compensates the propulsion speed.


\begin{acknowledgments}
  We gratefully acknowledge the DFG for funding within the priority program SPP 1726 (grant number SP 1382/3-2). AC has been supported by a DAAD WISE fellowship. AF acknowledges funding by the DFG through the Graduate School of Excellence ``Materials Science in Mainz'' (GSC 266). Numerical computations were carried out on the Mogon2 Cluster at ZDV Mainz.
\end{acknowledgments}

\appendix

\section{Hard-disc spinodal}
\label{sec:spin}

For completeness, here we provide the approximate solution of the condition $\al\gam+\Dr=0$ for the spinodal of hard discs. We define $\rho_0\equiv a\bar\rho_\text{c,ABP}=\frac{3}{4}$ and
\begin{equation*}
  z \equiv \frac{\rho_0(1+\chi)}{1+\rho_0\chi}\left(\frac{\bar\rho}{\bar\rho_\text{c,ABP}} - 1\right),
\end{equation*}
for which the solution reads
\begin{equation*}
  z_\pm = -\frac{1}{4}\rho_0^2\chi \pm \frac{1}{4}\sqrt{\rho_0^4\chi^2+1+2\rho_0^2\chi-\frac{16\Dr}{v_0^2}(1+\rho_0\chi)^3}.
\end{equation*}

\section{Soft-disc spinodal for constant speeds}
\label{sec:soft}

\begin{figure}[t]
  \centering
  \includegraphics[scale=.9]{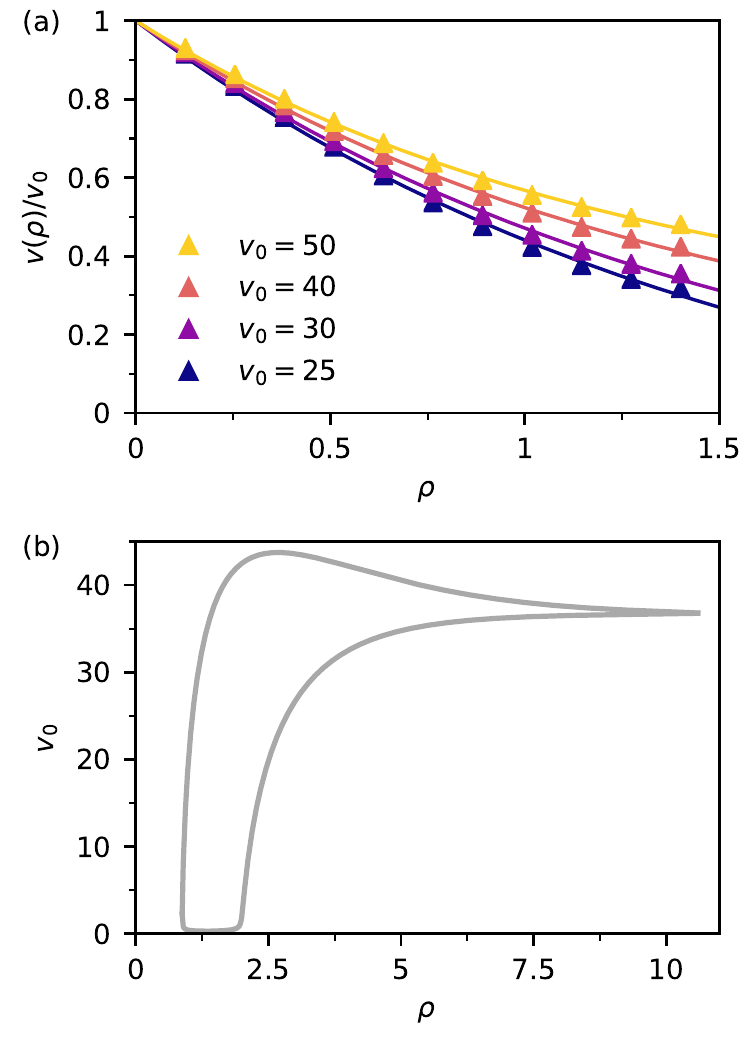}
  \caption{(a)~Effective speed $v(\rho)/v_0$ for soft disks at different $v_0$. The lines are fits of Eq.~\eqref{eq:veff:harm}. (b)~Mean-field spinodal enclosing the linearly unstable region. While the spinodal comes close to the lower axis, it remains nonzero so that for exactly $v_0=0$ the system is homogeneous.}
  \label{fig:harm:abp}
\end{figure}

In Fig.~\ref{fig:harm}, we have observed that for soft discs interacting through the harmonic potential Eq.~\eqref{eq:hp}, the binodals approach each other at larger speeds, implying reentrance into the disordered phase and the existence of an upper critical point. This behavior is quite different from the usual scenario found for ABPs, where the two phase region extends to infinite speeds. Here we briefly test whether a closed two-phase region can be reproduced, at least in principle, through a modified effective speed $v(\rho)$. To this end, we measure for pure ABPs the effective speed in the homogeneous region. The resulting reduced speeds $v(\rho)/v_0$ are plotted in Fig.~\ref{fig:harm:abp}(a) as a function of density $\rho$. In contrast to hard discs, we see that these speeds do not collapse onto a single linear decay, but curve away and at high densities depend on $v_0$. This non-linear slowdown reflects the ability of the soft particles to overlap and pass through each other, thus avoiding dynamic arrest at high densities for sufficiently high propulsion speeds.

We capture this behavior through the expression
\begin{equation}
  \label{eq:veff:harm}
  v_\text{HAR}(\rho) = v_0\left[(1-a)e^{-b\rho}+a\right],
\end{equation}
which fits the simulation data very well. For the coefficients $a$ and $b$ we find a linear dependence on $v_0$,
\begin{equation}
  a \simeq 0.017v_0-0.62, \quad b \simeq 0.006v_0+0.48.
\end{equation}
For $a>0$ (\emph{i.e.}, $v_0>36$), the effective speed would reach a non-vanishing value even as $\rho\to\infty$, thus avoiding dynamic arrest.

Following the derivation for hard disks in Sec.~\ref{sec:meanfield}, we solve the criterion $\al\gam+\Dr=0$ for the instability line numerically. The result is plotted in Fig.~\ref{fig:harm:abp}(b), which indeed describes a closed instability region in qualitative agreement with the behavior observed in the simulations. However, for large densities the shape of the instability region becomes somewhat implausible. Presumably this is an artifact of employing the fits Eq.~\eqref{eq:veff:harm} at high densities.


%

\end{document}